\documentclass[10pt]{article}
\usepackage[dvips]{graphicx,psfrag}
\usepackage{amssymb}
\usepackage{color}
\input epsf
\definecolor{Blue}{rgb}{0.3,0.3,0.9}
\definecolor{Red}{rgb}{1,0,0}
\definecolor{Green}{rgb}{0,1,0}
\newcommand{\be}{\begin{equation}}
\newcommand{\ee}{\end{equation}}
\newcommand{\bea}{\begin{eqnarray}}
\newcommand{\eea}{\end{eqnarray}}

\begin{document}
\begin{center}
{\Large\bf Physics in the multiverse: an introductory review}

\bigskip \bigskip \medskip
{\large Aur\'elien Barrau
}\\[.5cm]
{\it Laboratory for Subatomic Physics and Cosmology, Universit\'e Joseph Fourier,
CNRS/IN2P3\\ 53, avenue de Martyrs, 38026 Grenoble cedex, France\\
email: Aurelien.Barrau@cern.ch}\\[1cm]

{\bf Abstract}
\end{center}
This brief note, written for non-specialists, aims at drawing an introductive overview of the
multiverse issue.\\

{\it Published in:} CERN Courier, vol. 47, issue 10 (2007) pp. 13-17
\\[1cm]

Is our entire universe a tiny island within an infinitely vast and infinitely diversified meta-world ? This could be either one of the most important revolutions in the history of cosmogonies or merely a misleading statement that reflects our lack of understanding of the most fundamental laws of physics.

The idea in itself is far from being new: from Anaximander to David Lewis, philosophers have exhaustively considered this eventuality. But what is especially interesting today is that it emerges, almost naturally, from some of our best - but often most speculative - physical theories. The multiverse is no longer a model; it is a consequence of our models. It offers an obvious understanding of the strangeness of the physical state of our universe. The proposal is attractive and credible but it requires a profound rethinking of current physics.

At first glance, the multiverse seems to lie outside science because it cannot
be observed. How, following the prescription of Karl Popper, can a theory be
falsifiable if we cannot observe its predictions? This way of thinking is not
really correct for the multiverse for several reasons. First, predictions can be
made in the multiverse: it leads only to statistical results but this is also
true for any physical theory within our universe, owing both to fundamental
quantum fluctuations and to measurement uncertainties. Secondly, it has never
been necessary to check all the predictions of a theory to consider it as
legitimate science. General relativity, for example, has been extensively tested
in the visible world and this allows us to use it within black holes even though
it is not possible to go there to check. Finally, the critical rationalism of
Popper is not the final word in the philosophy of science. Sociologists,
aestheticians and epistemologists have shown that there are other demarcation
criteria to consider. History reminds us that the definition of science can only
come from within and from the praxis: no active area of intellectual creation
can be strictly delimited from outside. If scientists need to change the borders
of their own field of research, it would be hard to justify a philosophical
prescription preventing them from doing so. It is the same with art: nearly all
artistic innovations of the 20th century have transgressed the definition of art
as would have been given by a 19th century aesthetician. Just as with science
and scientists, art is internally defined by artists.\\

For all these reasons, it is worth considering seriously the possibility that we live in a multiverse. This could allow understanding of the two problems of  complexity and naturalness. The fact that the laws and couplings of physics appear to be fine-tuned to such an extent that life can exist and most fundamental quantities assume extremely "improbable" values would appear obvious if our entire universe were just a tiny part of a huge multiverse where different regions exhibit different laws. In this view, we are living in one of the "anthropically favoured" regions. This anthropic selection has strictly no teleological and no theological dimension and absolutely no link with any kind of "intelligent design": it is nothing other than the obvious generalization of the selection effect that already has to be taken into account within our own universe. When dealing with a sample, it is not possible to avoid wondering if it accurately represents the full set and this question must of course be asked when considering our universe within the multiverse.

The multiverse is not a theory. It appears as a consequence of some theories and
these have other predictions that can be tested within our own universe. There
are many different kinds of possible multiverses, depending on the particular
theories, some of them even being possibly interwoven.\\

The most elementary multiverse is simply the infinite space predicted by general relativity, at least for flat and hyperbolic geometries. An infinite number of Hubble volumes should fill this meta-world. In such a situation, everything that is possible (i.e. compatible with the laws of physics as we know them) should occur. This is so just because an event with a non-vanishing probability has to happen somewhere if space is infinite. The structure of the laws of physics and the values of fundamental parameters cannot be explained by this multiverse, but many specific circumstances can be understood by anthropic selections: some places are, for example, less homogenous than our Hubble volume, so we cannot live there because they are less life-friendly than our universe, where the primordial fluctuations are perfectly adapted as the seeds for structure formation.

General relativity also faces the multiverse issue when dealing with black holes. The maximal analytic extension of the Schwarzschild geometry, as exhibited by conformal Penrose-Carter diagrams, shows that another universe could be seen from within a black hole. This interesting feature is well known to disappear when the collapse is considered dynamically. The situation is however more interesting for charged or rotating black holes, where an infinite set of universes with attractive and repulsive gravity appear in the conformal diagram. The wormholes that possibly connect these universes are extremely unstable but this does not alter the fact that this solution reveals other universes (or other parts of our own universe, depending on the topology), whether accessible or not. This multiverse is however extremely speculative as it could very well be just a mathematical ghost. Furthermore, nothing allows us to understand explicitly how it formed.

A much more interesting pluriverse is associated with the interior of black holes when quantum corrections to general relativity are taken into account. Bounces should replace singularities in most quantum gravity approaches, and this leads to an expanding region of spacetime inside the black hole that can be considered as a universe. In this model our own universe would have been created by such a process and should also have a very large number of children universes, thanks to its numerous stellar and supermassive black holes. This could lead to a kind of cosmological natural selection where the laws of physics tend to maximize the number of black holes (just because such universes generate more universes of the same kind). It also allows for several possible observational tests that could refute the theory and does not rely on the use of any anthropic argument. However, it is not clear how the constants of physics could be inherited from the parent universe by the child universe with small random variations and the detailed model associated with this scenario does not yet exist.     

One of the richest multiverses is associated with the fascinating meeting of
inflationary cosmology and string theory. On the one hand, eternal inflation can
be understood by considering a massive scalar field. The field will have quantum
fluctuations, which will, in half of the regions, increase its value; in the
other half, the fluctuations will decrease the value of the field. In the half
where the field jumps up, the extra energy density will cause the universe to
expand faster than in the half where the field jumps down. After some time, more
than half the regions will have the higher value of the field, because they
simply expand faster than the low-field regions. The volume-averaged value of
the field will therefore rise and there will always be regions in which the
field is high: the inflation becomes eternal. The regions in which the scalar
field fluctuates downward will branch off from the eternally inflating tree and
exit inflation. On the other hand, string theory has recently faced a third
change of paradigm. After the revolutions of supersymmetry and duality, the
"landscape" has recently been discovered. This metaphoric word refers to the
very large number (maybe $10^{500}$) of possible false vacua of the theory. The known laws
of physics would just correspond to a specific island among many others. The
huge number of possibilities arises from different choices of Calabi-Yau
manifolds and different values of generalized magnetic fluxes over different
homology cycles. Among other enigmas, the incredibly strange value of the
cosmological constant (why are the 119 first decimals of the "natural" value
exactly compensated by some mysterious phenomena but not the 120th) would simply
appear as an anthropic selection effect within a multiverse where nearly every
possible value is realized somewhere. At this stage, every bubble-universe is
associated with one realization of the laws of physics and contains itself an
infinite space where all contingent phenomena take place somewhere. Because the
bubbles are causally disconnected forever (owing to the fast "space creation" by
inflation) it will not be possible to travel and discover new laws of physics. This multiverse - if true - would force a profound change of our deep understanding of physics. The laws reappear as kinds of phenomena; the ontological primer of our universe would have to be abandoned. At other places in the multiverse, there would be other laws, other constants, other numbers of dimensions; our world would be just a tiny sample. It could be, following Copernicus, Darwin and Freud, the fourth narcissistic injury.

Quantum mechanics was probably among the first branches of physics leading to
the idea of a multiverse. In some situations, it inevitably predicts
superposition. In order to avoid the existence of macroscopic Schrödinger cats
simultaneously living and dying, Bohr introduced a reduction postulate. This has
two considerable drawbacks: first it leads to an extremely intricate
philosophical interpretation where the correspondence between the mathematics
underlying the physical theory and the real world is no longer isomorphic (at
least not at any time), and, secondly, it violates unitarity. No known physical
phenomenon - not even the evaporation of black holes in its modern descriptions
- does this. These are good reasons for considering seriously the many-worlds interpretation of Hugh Everett. Every possible outcome to every event is allowed to define or exist in its own history or universe, via quantum decoherence instead of wavefunction collapse. In other words, there is a world where the cat is dead and another one where it is alive. This is simply a way of  trusting strictly the fundamental equations of quantum mechanics. The worlds are not spatially separated but more kinds of "parallel" universes. This tantalizing interpretation solves some paradoxes of quantum mechanics but remains very vague about how to determine when splitting of universes happens. This multiverse is complex and, depending on the very quantum nature of phenomena leading to other kind of multiverses, it could lead to higher or lower levels of diversity.

More speculative multiverses can also be imagined, associated with a kind of
platonic mathematical democracy or with nominalist relativism. In any case, it
is important to underline that the multiverse is not a hypothesis invented to
answer a specific question. It is simply a consequence of a theory usually built
for another purpose. Interestingly, this consequence also solves many complexity
and naturalness problems. In most cases, it even seems that the existence of
many worlds is closer to Ockham's razor (the principle of simplicity) than the
ad hoc assumptions that would have to be added to models to avoid the existence
of other universes.\\

Given a model, for example the string-inflation paradigm, is it possible to make
predictions in the multiverse? In principle, it is, at least in a Bayesian
approach. The probability  of observing vacuum $i$ (and the associated laws of
physics) is simply $P_i=P_i^{prior}f_i$ where $P_i^{prior}$ is determined by the geography of the landscape of
string theory and the dynamics of eternal inflation, and the selection factor 
$f_i$ characterizes the chances for an observer to evolve in vacuum $i$. This
distribution gives the probability for a randomly selected observer to be in a
given vacuum. Clearly, predictions can only be made probabilistically, but this
is already true in standard physics. The fact that we can observe only one
sample (our own universe) does not change the method qualitatively and still
allows the refuting of models at given confidence levels. The key points here
are the well known peculiarities of cosmology, even with only one universe: the
observer is embedded within the system described; the initial conditions are
critical; the experiment is "locally" irreproducible; the energies involved have
not been experimentally probed on Earth; and the arrow of time must be
conceptually reversed.

However this statistical approach to testing the multiverse suffers from severe
technical shortcomings. First, while it seems natural to identify the prior
probability with the fraction of volume occupied by a given vacuum, the result
depends sensitively on the choice of a space-like hypersurface on which the
distribution is to be evaluated. This is the so-called measure problem in the
multiverse. Secondly, it is impossible to give any sensible estimate of . This
would require an understanding of what life is - and even of what consciousness
is - and that simply remains out of reach for the time being. Except in some
favourable cases -- for example when all the universes of the multiverse present
a given characteristic that is incompatible with our universe -- it is very hard
to refute explicitly a model in the multiverse. But difficult in practice does
not mean intrinsically impossible. The multiverse remains within the realm of
Popperian science. It is not qualitatively different from other proposals
associated with usual ways of doing physics. Clearly, new mathematical tools and
far more accurate predictions in the landscape (which is basically totally
unknown) are needed for falsifiability to be more than an abstract principle in
this context. Moreover, falsifiability is just one criterion among many possible
ones and it should probably not be overdetermined.\\

When facing the question of the incredible fine-tuning required for the
fundamental parameters of physics to allow the emergence of complexity, there
are not many possible ways of thinking. If one does not want to use God or rely
on an unbelievable luck that led to extremely specific initial conditions, there
are mainly two remaining possible hypotheses. The first would be to consider
that since complexity - and in particular life - is an adaptive process, it
would have emerged in nearly any kind of universe. This is a tantalizing answer
but our own universe shows that life requires extremely specific conditions to
exist. It is hard to imagine life in a universe without chemistry, maybe without
bound states or with other numbers of dimensions. The second idea is to accept
the existence of many universes with different laws where we naturally find
ourselves in one of those compatible with complexity. The multiverse was not
imagined to answer this specific question but appears "spontaneously" in serious
physical theories, so it can be considered as the simplest explanation to the
puzzling issue of naturalness. This of course does not prove the model to be
correct, but it should be emphasised that there is absolutely no
"pre-Copernican" anthropocentrism in this thought process.\\

It could very well be that the whole idea of multiple universes is misleading. 
It could very well be that the discovery of the most fundamental laws of 
physics will make those parallel worlds totally obsolete in a few years. It 
could very well be that with the multiverse science is just entering a "no 
through road". Prudence is mandatory when physics tells us about invisible 
spaces. But it could also very well be that we are facing a deep change of 
paradigm that revolutionizes our understanding of nature and opens new fields 
of possible scientific thought. Because they lie on the border of science, 
these models are dangerous, but they offer the extraordinary possibility of 
constructive interference with other kinds of human knowledge. The multiverse 
is a risky thought -- but discovering new worlds has always been risky.


\begin{thebibliography}{07}
\bibitem{a1} A. Aguirre, in {\it Universe of Multiverse}, ed. B. Carr
(Cambridge: Cambridge University Press, 2007) pp. 367-386
\bibitem{a2} A. Barrau, in Proc. of the 11$^{th}$ 
Marcel Grossmann Meeting on General Relativity, eds.  Kleinert, R.T. Jantzen \&
R. Ruffini (Singapore: World Scientific, 2007), arXiv:gr-qc/0612045v2
\bibitem{b1} R. Bousso \& J. Polchinski, JHEP 0006 (2000) 006
\bibitem{b3} R. Bousso \& B. Freivogel, JHEP 0706 (2007) 018
\bibitem{b2} J.D. Bjorken, in {\it Universe of Multiverse}, ed. B. Carr
(Cambridge: Cambridge University Press, 2007) pp. 181-189
\bibitem{c1} B. Carter, Phil. Trans. Roy. Soc. A 310 (1983) 347
\bibitem{c2} B. Carter, Int. J. Theor. Phys. 43 (2004) 721
\bibitem{d1} P. Davies, Mod. Phys. Lett. A (2004) 727
\bibitem{e2} G. Ellis, in {\it Universe of Multiverse}, ed. B. Carr
(Cambridge: Cambridge University Press, 2007) pp. 387-409
\bibitem{e1} H. Everett, Rev. Mod. Phys. 29 (1957) 454
\bibitem{f1} B. Freivogel \& L. Susskind, Phys. Rev. D 70 (2004) 126007
\bibitem{h2} J.B. Hartle, in {\it Universe of Multiverse}, ed. B. Carr
(Cambridge: Cambridge University Press, 2007) pp. 275-284
\bibitem{h1} S. Hawking, in {\it Universe of Multiverse}, ed. B. Carr
(Cambridge: Cambridge University Press, 2007) pp. 91-98
\bibitem{k1} S. Krachru, R. Kallosh, A.D. Linde \& S.T. Trivedi, Phys. Rev. D
68 (2003) 046005
\bibitem{l1} A.D. Linde, Phys. Lett. B 129 (1983) 177
\bibitem{l4} A.D. Linde, in {\it The Very Early Universe}, eds. G.W. Gibbons,
S.W. Hawking \& S. Siklos (Cambridge: Cambridge University Press, 1983)
\bibitem{l2} A.D. Linde, Phys. Lett. B 175 (1986) 2848
\bibitem{l5} A.D. Linde, Rep. Prog. Phys. 47 (1984) 194
\bibitem{l3} A.D. Linde, Phys. Script. T 15 (1987) 100
\bibitem{l8} A.D. Linde, {\it Particle Physics and Inflationary Cosmology}
(Chur, Switzerland: Harwood Academic, 1990)
\bibitem{l6} A.D. Linde, Sci. Am. 271 (1994) 32
\bibitem{l7} A.D. Linde, in {\it Universe of Multiverse}, ed. B. Carr
(Cambridge: Cambridge University Press, 2007) pp. 127-149
\bibitem{m1} V. Mukhanov, in {\it Universe of Multiverse}, ed. B. Carr
(Cambridge: Cambridge University Press, 2007) pp. 267-273
\bibitem{r1} M. Rees, in {\it Universe of Multiverse}, ed. B. Carr
(Cambridge: Cambridge University Press, 2007) pp. 57-75
\bibitem{s2} L. Smolin, in {\it Universe of Multiverse}, ed. B. Carr
(Cambridge: Cambridge University Press, 2007) pp. 323-366
\bibitem{s3} L. Smolin, {\it The Life of the Cosmos} (New York: Oxford
University Press, 1997)
\bibitem{s4} L. Smolin, Physica A 340 (2004) 705
\bibitem{s5} L. Smolin, arXiv:hep-th/0612185v1 (2006)
\bibitem{s3} L. Susskind, arXiv:hep-th/0405189v3 (2003)
\bibitem{s2} L. Susskind, {\it The cosmic landscape: String theory and the 
illusion of intelligent design.}
(New York, USA: Little, Brown, 2005)
\bibitem{s1} L. Susskind, in {\it Universe of Multiverse}, ed. B. Carr
(Cambridge: Cambridge University Press, 2007) pp. 247-266
\bibitem{teg4} M. Tegmark, Class. Quantum Grav. 14 (1997) L69
\bibitem{teg2} M. Tegmark, Ann. Phys. 270 (1998) 1
\bibitem{teg3} M. Tegmark, JCAP 0504 (2004) 001
\bibitem{teg1} M. Tegmark, in {\it Universe of Multiverse}, ed. B. Carr
(Cambridge: Cambridge University Press, 2007) pp. 99-125
\bibitem{v1} A. Vilenkin, Phys. Lett. D 27 (1983) 2848
\bibitem{v2} A. Vilenkin, in {\it Universe of Multiverse}, ed. B. Carr
(Cambridge: Cambridge University Press, 2007) pp. 163-179
\bibitem{v3} A. Vilenkin, J. Phys. A 40 (2007) 6777
\bibitem{w2} S. Weinberg, Phys. Rev. Lett. 59 (1987) 2607
\bibitem{w1} S. Weinberg, in {\it Universe of Multiverse}, ed. B. Carr
(Cambridge: Cambridge University Press, 2007) pp. 29-42
\bibitem{w3} F. Wilczek, in {\it Universe of Multiverse}, ed. B. Carr
(Cambridge: Cambridge University Press, 2007) pp. 43-54
\end{thebibliography}
\end{document}